# Interrelation of epitaxial strain and oxygen deficiency in La$_{0.7}$Ca$_{0.3}$MnO$_{3-\delta}$ thin films

K. Kawashima, G. Logvenov, G. Christiani and H.-U. Habermeier*

Max-Planck-Institute for Solid State Research

Heisenbergstr. 1     D 50569 Stuttgart, Germany

**Abstract**
The interrelation between the epitaxial strain and oxygen deficiency in La$_{0.7}$Ca$_{0.3}$MnO$_{3-\delta}$ thin films was studied in terms of structural and functional properties. The films with a thickness of 1000Å were prepared using a PLD system equipped with a RHEED facility and a pyrometric film temperature control. The epitaxial strain and the oxygen deficiency in the samples were systematically modified using three different substrates: SrTiO$_3$, (LaAlO$_3$)$_{0.3}$-(Sr$_2$AlTaO$_6$)$_{0.7}$ and LaSrAlO$_4$, and four different oxygen pressures during film growth ranging from 0.27mbar to 0.1mbar. It could be demonstrated that the oxygen incorporation depends on the epitaxial strain: oxygen vacancies were induced to accommodate tensile strain whereas the compressive strain suppressed the generation of oxygen vacancies.

*Corresponding author: H.-U. Habermeier e-mail: huh@fkf.mpg.de

# 1. Introduction

Infinite layer manganites of the type $AMnO_3$, where the A site can be occupied by a rare earth metal, an alkaline earth metal or a combination of both, exhibit fascinating properties because of their rich phase diagram including high spin polarization of the conducting electrons, the colossal magnetoresistance (CMR) effect [1-4] and the paramagnetic insulator to ferromagnetic metal transition [5]. These properties are consequences of the partial substitution of the three-valent rare earth ion by a divalent alkaline earth one. $La_{1-x}Ca_xMnO_{3-\delta}$ (LCMO) may serve as a prototype material. Strong electron correlation and a coupling of charge-, spin-, and orbital degrees of freedom superimposed by lattice interactions determine their properties. In the family of the rare earth manganites, these interactions are governed by two dominant mechanisms. One is the double exchange (DE) mechanism within the Mn-O-Mn building block [6], which causes ferromagnetic spin alignment and enhances the conductivity, and the other one is based on the Jahn-Teller (JT) effect due to the $Mn^{3+}$ ions inducing a structural reconstruction to accommodate the JT distortions and resulting in a removal of the $e_g$ and $t_{2g}$ orbital degeneracy. The magnitudes of both effects are sensitive to the geometrical configuration of the $O_6$ octahedra surrounding the Mn ions and their competition gives rise to an intriguing interplay of different electronic and magnetic phases which can be tuned extrinsically, especially in thin films. Due to scientific interests for its remarkable properties such as the large CMR effect and the high spin-polarization of the conducting electrons as well as their application potential in spintronic devices, intensive studies were devoted to a fundamental understanding of the nature of the electronic and magnetic properties of LCMO thin films. One of the main topics in the research is the effect of the epitaxial strain on their electronic properties. Using an appropriate substrate it is possible to induce a tensile or compressive strain to epitaxially grown thin films. When the LCMO lattice is strained, the $MnO_6$ octahedra tend to distort and/or rotate to accommodate it, and consequently Mn-O bond lengths and Mn-O-Mn bond angles are modified [7,8]. These modifications affect the competition between the DE interaction and JT distortions, and lead to a change in the transport and magnetic properties [9-12]. To use the strain is regarded as a straightforward and powerful method to develop the understanding of the nature of the properties of LCMO thin films. Additionally, the oxygen stoichiometry plays an important role in determining the properties of LCMO. Oxygen vacancies drive the Mn valence state from $Mn^{4+}$ to $Mn^{3+}$, and the functional properties will be modified accordingly [13]. In general, the concentration of oxygen vacancies is determined by the growth thermodynamics, growth kinetics, and post-annealing. Several papers dealing with oxygen incorporation and depletion [14, 15] in manganite thin films have already been published. Lattice strain and the oxygen deficiency are considered as the key sources to tune the properties of the rare earth manganite thin films for a given cation doping level, both, electronically and geometrically.

We would like to note that the lattice strain and the strain caused by oxygen vacancies are coupled in the real metal oxide systems and each of their contributions cannot be separated. It was pointed out that the lattice mismatch between film and substrate can impact the oxygen incorporation [16, 17] via minimizing the elastic energy. In this paper we report a systematic study using LCMO thin films to shed light on this interrelation and develop a general view. We intentionally changed the lattice strain and the oxygen concentration in our samples. The lattice strain was modified by using different substrates: $SrTiO_3$ (STO), $(LaAlO_3)_{0.3}$-$(Sr_2AlTaO_6)_{0.7}$ (LSAT) and $LaSrAlO_4$ (LSAO), and the concentration of oxygen vacancies was altered by using four different oxygen pressures during growth, namely, 0.27, 0.2, 0.13 and 0.1mbar [18]. Structural and functional measurements were performed on all samples, and the changes of the properties were discussed in terms of the lattice strain and the oxygen deficiency.

## 2. Experiments

Doped manganites thin films $La_{1-x}Ca_xMnO_{3-\delta}$ with x=0.3 were grown by the pulsed laser deposition (PLD) method. For the PLD process a KrF excimer laser with a wavelength of λ=248nm was used, applying a commercial LCMO target (Lesker) and an energy fluency at the target of 1.5 J/cm². We grew 1000Å LCMO films on (001) STO, (001) LSAO and (001) LSAT substrates. All substrates have been subjected to an acetone and isopropanol ultrasonic cleaning prior to loading them into the PLD growth chamber. The room temperature in-plane lattice parameter of pseudo-cubic LCMO is 3.86Å. The lattice constants of STO, LSAO and LSAT substrates are 3.905, 3.75 and 3.87Å, respectively. The lattice mismatch, f, between the LCMO films and the substrates, calculated as $f = (a_l-a_s)/a_s$, are -1.2, 2.9 and -0.3%, for STO, LSAO and LSAT, respectively, where $a_l$ and $a_s$ are the in-plane lattice parameters of the unstrained LCMO layers and the substrates. Therefore the LCMO films deposited on STO substrates are under tensile strain. In contrast, the LCMO films on LSAO substrates are under compressive strain. The lattice mismatch between the LCMO and LSAT substrate is negligible. The deposition temperature was kept at 750°C in this work. After deposition, the samples were cooled down to 530°C with 15°C/min then the oxygen pressure was set to 1bar. The post-annealing lasted for 1 hour; afterwards the infrared laser heater for the substrate was turned off. The samples cooled down to room temperature within 10 min. For each substrate, four LCMO thin films were deposited with different oxygen pressures, namely, 0.27, 0.2, 0.13 and 0.1 mbar. The combination of two parameters, the controlled lattice mismatch and the oxygen pressure during film growth, enables us to explore the interrelation between the oxygen deficiency in the LCMO films and the lattice strain. After the LCMO film growth, the surface morphology was investigated by atomic force microscopy (AFM). X-ray diffraction (XRD) as well as reciprocal space mapping (RSM) was used to characterize the crystal structure and the epitaxial strain states. Finally, a Quantum Design system SQUID vibrating sample magnetometer (VSM) and a PPMS Quantum Design system were applied to investigate the magnetic and transport properties. For the transport measurements Au contacts for standard four contact methods were evaporated on the top of the samples.

## 3. Results and discussions

### 3.1 Surface morphology

Fig. 1(a) shows the dependence of the root mean square (RMS) surface roughness measured by AFM versus the oxygen pressure during growth of all our LCMO films. The samples which were deposited at pressures lower than 0.2 mbar showed RMS values less than 1 nm, a higher oxygen pressure (0.27mbar) causes an increase of the RMS values up to 2.7-4.2 nm. This tendency was observed for the different substrates. We conclude that increase of the oxygen pressure causes a crossover from a layer-by-layer growth mode to an island growth mode. This conjecture follows from the fact that the PLD deposition rate increases with decrease of the oxygen pressure for the same laser fluency. This observation is not due to a change of the surface mobility of the oncoming species. An increase of the deposition rate with decrease of the oxygen pressure causes the transition from a 3D island growth mode to a 2D layer-by-layer growth mode. Fig. 1(b) and (c) show the AFM images of LCMO films on LSAT grown at 0.27 mbar and 0.1 mbar, respectively. The LCMO film grown at 0.27 mbar oxygen pressure shows the morphology due to the island growth. On the other hand, the surface steps caused by layer-by-layer growth mode are observed on the LCMO film grown at 0.1 mbar. Similar surface morphology was obtained for the LCMO film grown on LSAO substrate at reduced oxygen pressure 0.1 mbar shown in Fig. 1(d).

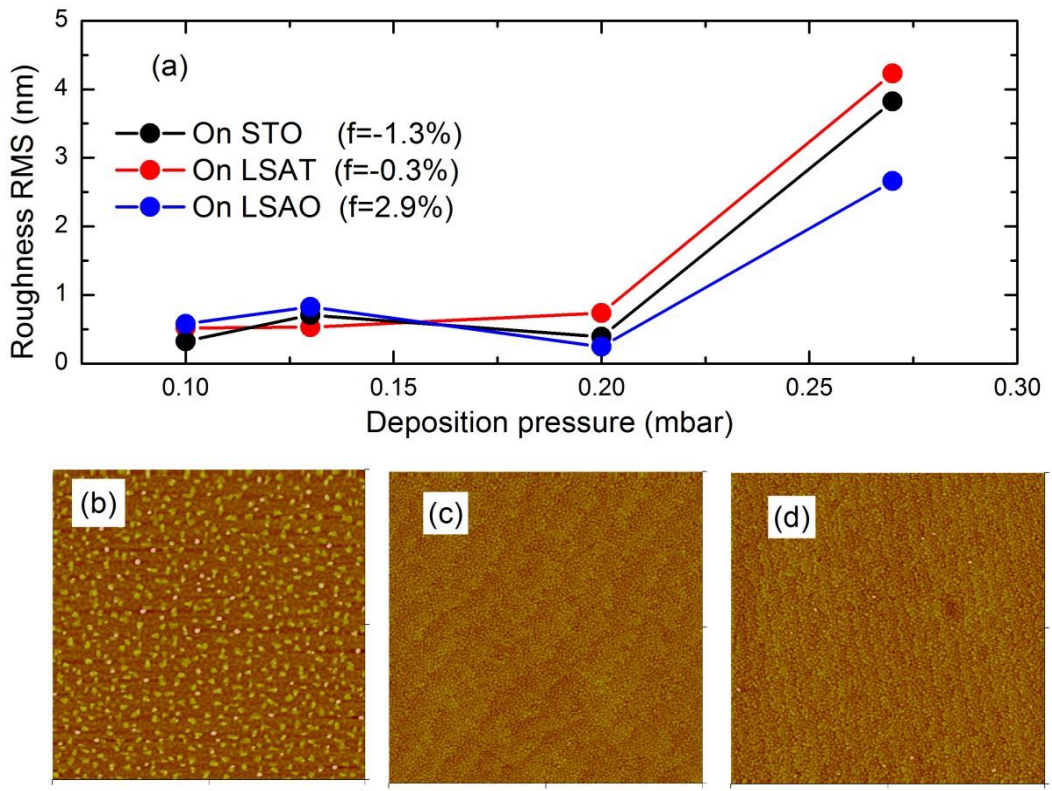

*Fig.1 (a) Relation between the surface roughness and the deposition pressures of all the samples. The AFM images of 1000A LCMO on LSAT deposited at 0.27mbar (b) and at 0.1mbar (c). The AFM image of LCMO on LSAO deposited at 0.1 mbar (d).*

## 3.2 XRD analysis

The results of the XRD θ-2θ scan are presented in Fig. 2. Here, the magnification around the LCMO (002) peak is shown. The diffraction intensity is normalized to the peak intensity of the substrates. For the LCMO film grown at 0.27 mbar oxygen pressure on STO substrate the (002) peak position is at higher angle than the bulk value. This corresponds to reduced out-of-plane lattice constant compared to the bulk value. For tensile strain the in-plane lattice constant stretches, while the out-of-plane lattice constant shrinks. For LCMO films deposited at reduced oxygen pressures the (002) peak positions shift to the lower angles, what corresponds to the increase of the out-of-plane lattice constant. The elongation of out-of-plane parameter indicates the increase of the oxygen deficiency in the film [13, 15, 19] since oxygen vacancies cause the increase of the Mn ionic radius due to change of the Mn valence state from $Mn^{4+}$ to $Mn^{3+}$. The ionic radius of $Mn^{3+}$ is 0.7Å as compared to 0.5Å for $Mn^{4+}$ [19]. The shift of the (002) film peak position confirms that the concentration of oxygen vacancies was modified by the change of the oxygen pressure during growth. The high quality of the crystal structure and the small interface roughness of the LCMO film on STO substrate deposited at low pressures can be judged from the finite-thickness oscillations. The presence or absence of finite-thickness fringes is a fingerprint of the quality of the two interfaces, the film/substrate and the top one, as well as the microstructure of the film. The LCMO film grown at 0.27 mbar oxygen pressure does not show the finite-thickness oscillations. This result is consistent with the fact that the surface of this film measured by AFM is rough compared to films grown at lower oxygen pressures.

The LCMO films grown on LSAT substrates with a small lattice mismatch also show the successive shift of the (002) peak position to lower angles due to generation of oxygen vacancies during film growth at low oxygen pressures. Since the LCMO (002) diffraction peak position of the sample deposited at 0.27 mbar oxygen pressure matches the calculated position 47.06° corresponding to the bulk out-of-plane lattice constant 3.86Å, this LCMO film is least affected by the strain as expected due to the small lattice mismatch. Here, again finite-thickness oscillations peaks ( albeit less pronounced as compared to the STO case ) are visible on the right side of the main peak, except for the sample deposited at 0.27 mbar oxygen pressure, where the film surface is relatively rough.

In LCMO films grown on LSAO substrates, the position of the LCMO (002) peak is at a lower angle (see Fig. 2(c)) compared to the position expected for the bulk, this corresponds to a larger out-of-plane lattice constant compared to the bulk one. The corresponding (006) substrate peak of LSAO at $2\theta = 42.91^0$ is out of the scale if Fig. 2c. This elongation of the out-of-plane constant could be caused by the compressive strain. In these samples, the shift of the (002) peak position due to the change of the oxygen pressure during growth is not noticeable in contrast to the cases of STO and LSAT substrates. Instead of that, there is a large change in the full width at half maximum (FWHM) of the thin film peaks as shown in Fig. 3. The LCMO films on LSAO substrates have larger FWHM values compared to the values of LCMO films grown on STO and LSAT substrates. The large values of FWHM of the LCMO film on LSAO substrate grown at 0.27 and 0.2 mbar is caused by the inhomogeneity of the out-of-plane lattice parameter and the mosaicity, most likely due to strain relaxation and island growth mode which is fostered by a compressive strain [12]. The vertical lines in Fig. 2 correspond to the expected positions of the LCMO (002) peak assuming entirely stoichiometric using a Poisson ratio of $v = 0.35$.

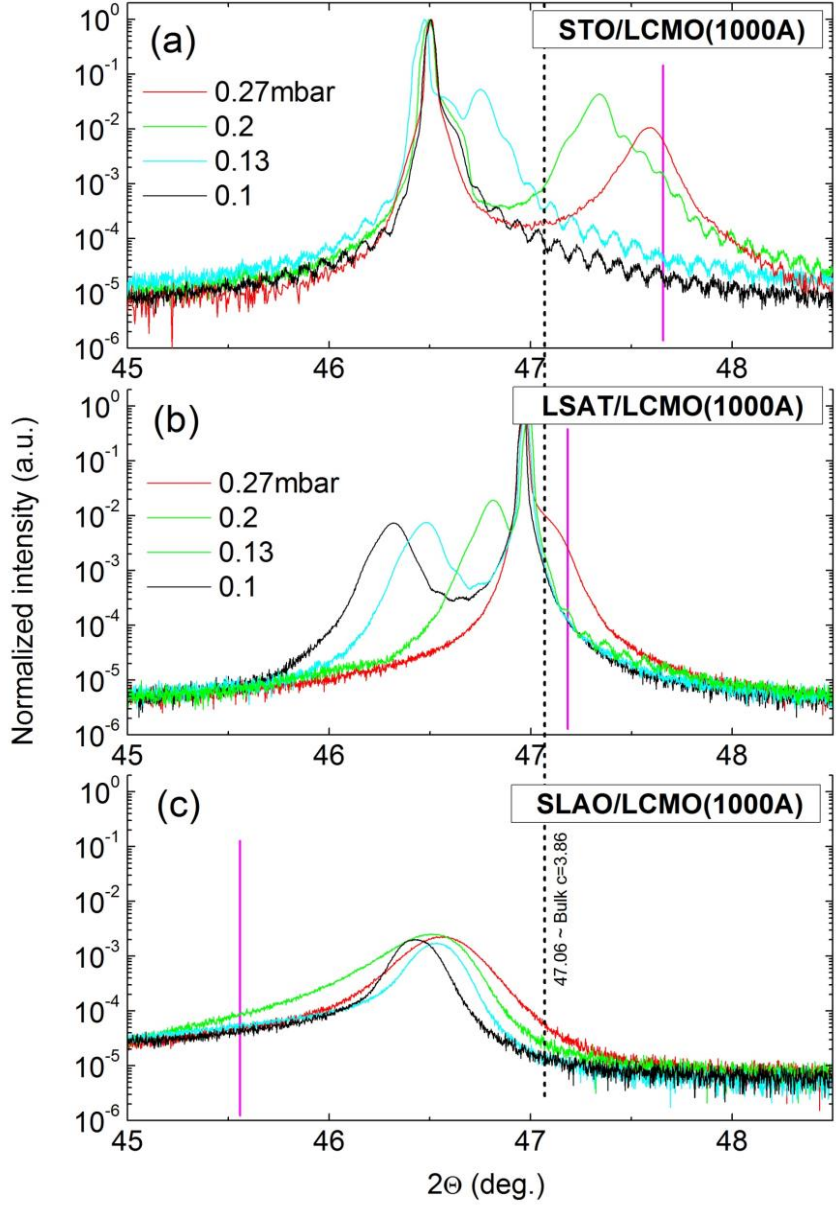

*Fig. 2 Fig.2 XRD scans around LCMO (002) peak. The diffraction intensity is normalized to STO(002), LSAT(002) and LSAO(006), respectively. The (006) diffraction peak of LSAO substrate is at 42.91°, which is out of the range shown in the figure. The vertical lines (magenta ) correspond to the expected positions of fully strained and entirely stoichiometric LCMO films, the dotted line ( black ) represents the position of the bulk LCMO (002) peak.*

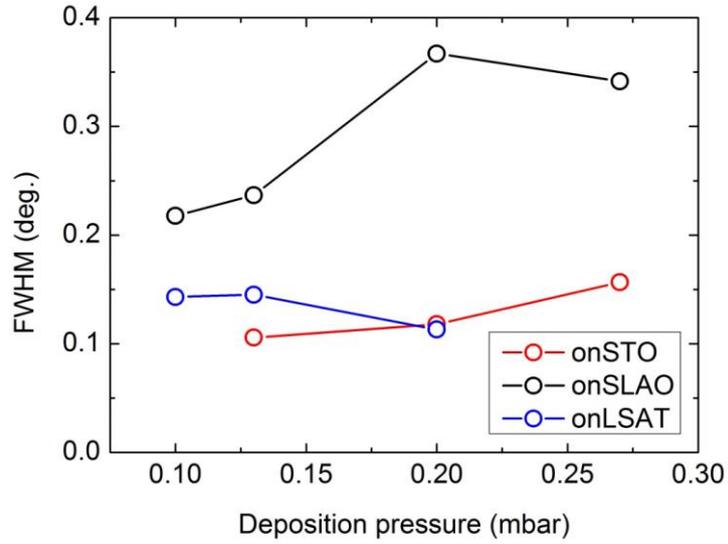

*Fig.3 Values of full width at half maximum of the LCMO (002) peak. In some samples, the values are missing since the LCMO peaks overlap with the peaks from the substrate.*

Further analysis of the crystal structure of our LCMO films by using reciprocal space map (RSM) plots supports this view. Fig. 4 shows the RSM plots of the diffracted X-ray intensity distribution in the vicinity of the (013) Bragg peaks of the LCMO films on STO and on LSAT substrates.

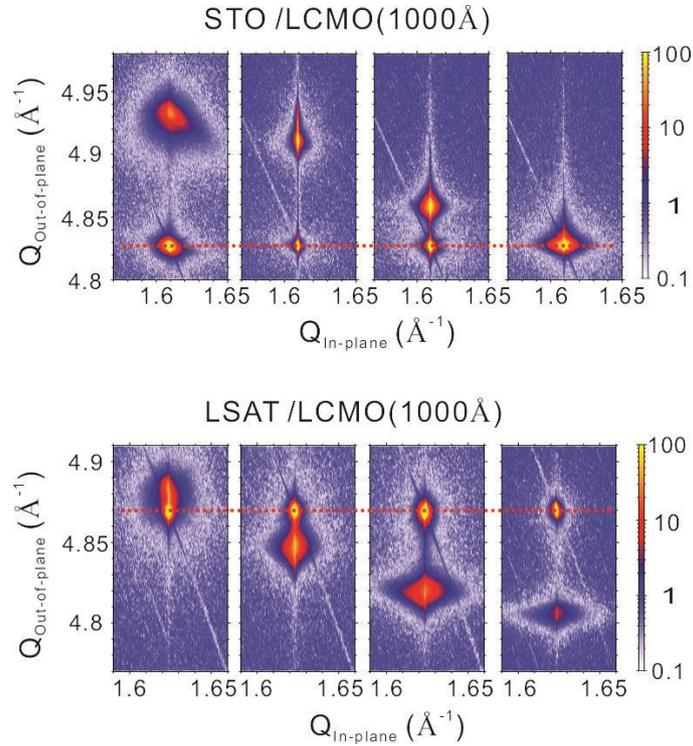

*Fig. 4 Reciprocal space maps (RSM). The figures are corresponding to the samples grown at 0.27, 0.2, 0.13, and 0.1mbar from the left to the right. (a-d) RSM of LCMO on STO around (013) STO diffraction.(e-h) RSM of LCMO on LSAT around (013) LSAT diffraction.*

All LCMO films on these substrates are strained to the substrate and the shift of the out-of-plane Q value is clearly visible. Note that the shift of the out-of-plane Q value corresponds to the change of out-of-plane lattice constant obtained from the θ-2θ XRD scans. In the LCMO film grown on STO substrate at 0.27 mbar oxygen pressure, the shape of the diffraction peak is distorted in an asymmetric way. This is related to a partial relaxation of the crystal structure in the film. The strain-free island formation takes place earlier in this film compared to the other samples grown at lower oxygen pressures. Fig. 5 shows RSM plots of the (013) LCMO Bragg peak for four films grown on LSAO substrates at different oxygen pressures.

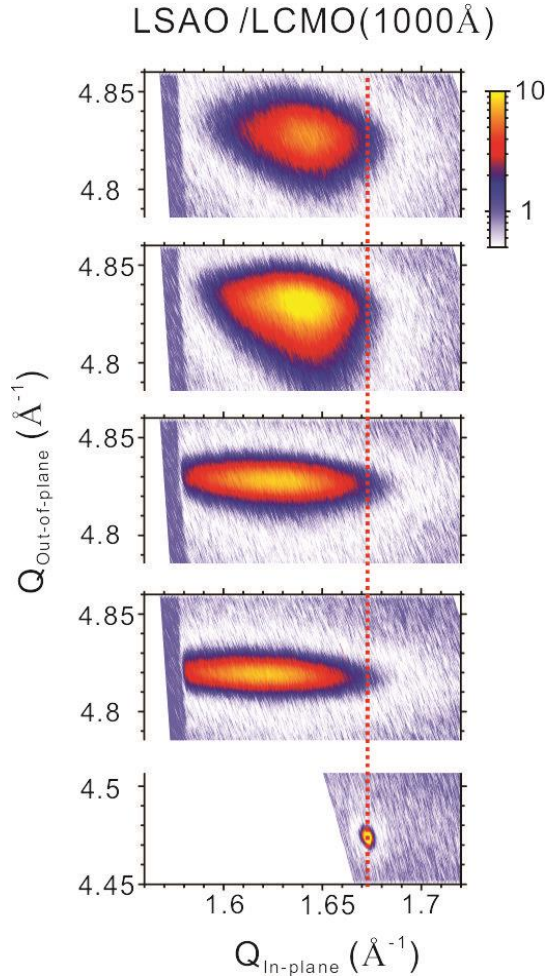

*Fig.5 RSM of LCMO film on LSAO substrate. The measurement was done around (019) LSAO reflection. The deposition pressures are, from the top to the bottom, (a) 0.27, (b) 0.2, (c) 0.13, (d) 0.1mbar; (e) corresponds to the (019) LSAO substrate peak.*

The in-plane and out-of-plane Q's for the (013) Bragg peak of bulk LCMO are $1.628 Å^{-1}$ and $4.883 Å^{-1}$, respectively, which are far from the (013) Bragg peak in our LCMO films on LSAO. Therefore those films are partially strained compressively and partially relaxed. The out-of-plane Q value of the (013) Bragg peak positions of the samples grown at different growth pressures shows tiny shift from $4.8291 Å^{-1}$ to $4.8188 Å^{-1}$, which is consistent to the small change of out-of plane lattice constant observed in θ-2θ scan. In addition, the shape of the RSM spot changes in the films grown at different oxygen pressures. Paying attention to the change of the shape of the (013) diffraction spots, the broadening of the out-of-plane Q value becomes smaller in the films deposited at lower oxygen pressures. This tendency coincides with the change of the FWHM of the (002) LCMO Bragg peak value presented in Fig. 3 and is related to the film microstructure. The (013) diffraction spots in the LCMO films deposited at 0.27 and 0.2 mbar oxygen pressures show broadening of both, the in-plane and out-of-plane Q values, indicating the relaxation due to the formation of stress-free or at least stress poor islands. The spots of the atomically smooth films deposited at 0.13 and 0.1 mbar oxygen pressures show a further broadening of the in-plane Q value, but shrinkage of the out-of-plane Q value. This observation indicates that these films have large spread in terms of the in-plane lattice constant but the out-of-plane parameter is more homogeneous. Taking into account the fact that the LCMO films grown on LSAO substrates at low oxygen pressure shows surface steps in the AFM image (shown in Fig.1 (d)) while films grown at higher oxygen pressure are rough, we conclude that change of the shape of the (013) Bragg spots is caused by the change of the LCMO film microstructure. The compressive strain is accommodated by creating twin domains, which has been observed not only in manganites [20,21] but also in perovskite $LaNiO_3$ [22] when they are in a compressive strain state.

### 3.3 Magnetism and transport measurements

In Fig. 6 the Curie temperature, $T_C$, and the peak temperature, $T_P$, are plotted as a function of the oxygen pressure during growth. $T_C$ is determined as the temperature where the derivative, dM/dT of the magnetization M measured at 3000 Oe parallel to the film plane in the field cooling (FC) regime shows the maximum. $T_P$ is determined from the transport measurement as the temperature where the resistivity has its maximum. The samples grown at oxygen pressures lower than 0.13 mbar show insulating behavior below 150 K, therefore no $T_P$ values can be determined for these films (see Fig. 6). The other physical properties such as saturation magnetization $M_S$, residual resistivity $\rho_0$ and coercive field $H_C$ are summarized in Table. I.

In the following we discuss the results of the transport and magnetic measurements for the films deposited on the different substrates and oxygen pressures with respect to the conjecture of the role of growth-induced oxygen vacancies. It must be stated, however, that the presence of oxygen vacancies implies a change of the Mn valency from $Mn^{4+}$ to $Mn^{3+}$ which cause in turn a shift of the overall doping level with implications to the resistivity, residual resistivity, the magnetic moment and $T_C$ as well.

The LCMO film grown on LSAT substrate - the one with the smallest lattice mismatch - at 0.27 mbar oxygen pressure has a saturation magnetization of 3.6 $\mu_B$ per Mn atom, in good agreement with the value, 3.7 $\mu_B$ per Mn atom for a bulk $La_{0.7}Ca_{0.3}MnO_3$. The residual resistivity, $\rho_0$, which is related to scattering of carriers due to defects, can be used as an indicator of the quality of the crystallinity [5]. It is 104 $\mu\Omega$cm which is similar to the values of the LCMO films reported in the literature [5, 23]. The coercive field, $H_C$, shows a quite small value of 57.6 Oe, indicating a weak pinning of the Bloch walls by intrinsic defects.

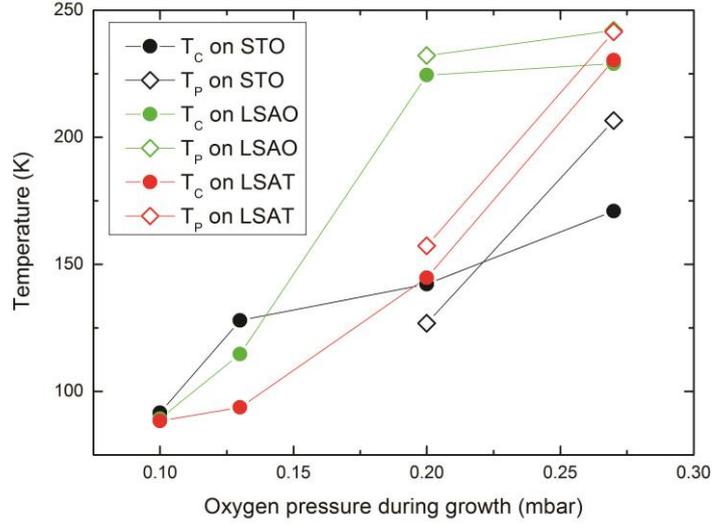

*Fig.6. Dependence of the Curie temperature $T_C$ and the peak temperature $T_P$ on the oxygen pressure during growth.*

### On STO

| $P_{Growth}$(mbar) | 0.27 | 0.2 | 0.13 | 0.1 |
|---|---|---|---|---|
| $M_S(\mu_B/Mn)$ | 3.3 | 3.0 | 2.3 | 1.2 |
| $\rho_0(\mu\Omega cm)$ | 411 | 5867 | — | — |
| $H_c$(Oe) | 254.4 | 357.3 | 286.0 | 547.2 |

### On LSAT

| $P_{Growth}$(mbar) | 0.27 | 0.2 | 0.13 | 0.1 |
|---|---|---|---|---|
| $M_S(\mu_B/Mn)$ | 3.6 | 2.9 | 1.1 | 0.9 |
| $\rho_0(\mu\Omega cm)$ | 104 | 4412 | — | — |
| $H_c$(Oe) | 57.6 | 114.3 | 375.0 | 341.0 |

### On LSAO

| $P_{Growth}$(mbar) | 0.27 | 0.2 | 0.13 | 0.1 |
|---|---|---|---|---|
| $M_S(\mu_B/Mn)$ | 3.2 | 2.7 | 1.5 | 0.9 |
| $\rho_0(\mu\Omega cm)$ | 161 | 118 | — | — |
| $H_c$(Oe) | 358.2 | 309.9 | 472.4 | 652.4 |

*Table. List of the physical properties such as saturation magnetization $M_S$, residual resistivity $\rho_0$, and coercive field $H_C$*

The tensile-strained LCMO film on STO substrate grown at 0.27 mbar oxygen pressure shows a reduced Curie temperature $T_C \approx 171K$ compared to LCMO films grown on LSAT and LSAO substrates at 0.27 mbar. We consider two possible scenarios for this reduction of $T_C$: (i) an extension of the Mn-O bond length due to tensile strain and (ii) the oxygen deficiency. According to the DE mechanism stretching of the Mn-O bonds lengths and the increase of the bonding angles due to a tensile strain suppresses the probability of electron hopping, then the ferromagnetism is weakened and $T_C$ is reduced [24]. The oxygen deficiency could also cause a reduction of $T_C$ since the oxygen vacancies in the crystal modifies the manganese valence from $Mn^{4+}$ to $Mn^{3+}$ by giving additional electrons. The reduction of $T_C$ due to the extension of the Mn-O bond lengths is observed in the thin LCMO films in which the effective tensile strain is strong, and the effect disappears in the thick films with the thickness more than 1000Å [25, 26]. Therefore, we attribute the reduction of $T_C$ in our LCMO films to the oxygen deficiency. In addition, the onset of the magnetic transition shown in Fig. 7(a) is at around 250 K in all LCMO films grown at oxygen pressure 0.27 mbar; however the magnetic transition is broad in tensile strained LCMO films on STO substrate compared with films grown on LSAT and LSAO substrates. Similar broadening of the magnetic transition is observed in the LCMO films grown at lower oxygen pressures. The transition broadening increases as the growth oxygen pressure reduces (see the case of LCMO film on LSAT substrate shown in Fig. 7(a)). Thus the LCMO film grown on STO substrate at 0.27 mbar oxygen pressure has certain concentration of the oxygen vacancies. Most likely the oxygen vacancies are induced to accommodate the tensile strain [17]. Relatively large residual resistivity 411 μΩcm in this film, which is 4 times larger than in the LCMO film on LSAT substrate, supports this conclusion. Indeed the oxygen vacancies are electronic defects and scatter charge carriers. The Curie temperature also reduces due to the presence of oxygen vacancies in this LCMO film under tensile strain.

The LCMO films grown at 0.2 mbar oxygen pressures on STO and LSAT substrates show reduced $T_C$ and $T_P$, and an increase of the residual resistance. The LCMO films grown at 0.13 and 0.1 mbar oxygen pressures on STO and LSAT substrates are insulating with significantly reduced magnetism. Taking into account that the out-of-plane lattice constant also increases as oxygen pressure decreases we attribute these tendencies to a rise of the concentration of the oxygen vacancies. As the oxygen deficiency is enhanced, the manganese valence changes and causes the reduction of magnetism accompanying with $T_C$ as well as $T_P$ and $M_s$. The residual resistivities also increase significantly.

The Curie temperature of the LCMO film on LSAO substrate grown at 0.2 mbar oxygen pressure does not change compared to the film grown at 0.27 mbar. The values of $\rho_0$ are 161 and 118μΩcm for the samples grown at 0.27 and 0.2mbar, respectively, and those are similar to the one of fully oxygenated LCMO grown on LSAT at 0.27mbar. This indicates that in the LCMO films grown on LSAO substrate the generation of oxygen vacancies is less favorable than in the films grown on STO and LSAT substrates. The compressive strain supports the oxidization of LCMO films. The magnetization hysteresis loops at 5K for LCMO grown at 0.27mbar on different substrates are shown in Fig. 7(b). Here, the slight discrepancy of the virgin curve and and the major hysteresis loops are regarded to be due to the residual magnetic field of the superconducting solenoid even after low temperature degaussing. The magnetic moment for film grown on LSAO substrate is not saturated at the field of 3000 Oe, where LCMO films on STO and LSAT substrates get saturated. This indicates that spins are strongly pinned due to structural distortion caused by a compressive strain.

LCMO films grown on LSAO substrate at 0.13 and 0.1 mbar show the insulating behavior without a characteristic peak of resistivity in the transport measurement. $T_C$'s are 115K and 89K, respectively, and the ferromagnetic moments are also reduced as shown in Table I. In the case of LCMO on LSAO, there are two possible reasons for the reduction of conductivity and ferromagnetism. One is the oxygen deficiency as is the case of LCMO on STO and LSAT. Although the out-of-plane lattice parameter does not show a noticeable change, there is a distinct spread in the in-plane parameter (Fig. 5). Therefore a certain amount of oxygen vacancies could be generated in the crystal system despite of the epitaxial compressive strain. The second possible reason is that the creation of twin domains disturbs electron hopping and suppresses the DE mechanism. As discussed in RSM of Fig. 5, the shape of (013) diffraction spot of the samples grown at 0.13 and 0.1mbar indicates a formation of twin domains, where the domains tilted with each other by an identical angle to accommodate a compressive strain. At the interfaces between tilted domains the connection between adjacent $MnO_6$ octahedra is distorted. And the DE interaction gets suppressed. Accordingly the conductivity and ferromagnetism are suppressed. It is also likely that both effects are combined, namely, the oxygen vacancies can be accommodated by forming twin domains. In this case, both of the oxygen vacancies and the distortion of octahedral contribute on the suppression of the conductivity and ferromagnetism.

### 4. Summary

We have grown LCMO thin films on three different substrates, STO, LSAT and LSAO, and used four different oxygen pressures for growth, 0.27, 0.2, 0.13 and 0.1mbar, in order to change the concentration of the oxygen vacancies. According to XRD analysis, the quality of crystal structure is good, and the oxygen deficiency becomes larger as the oxygen growth pressure becomes smaller, enabling us to perform a systematic study on the interrelation between strain and oxygen deficiency. The shift of the (002) X-ray diffraction peak due to the change of the oxygen growth pressure is large in the film with tensile strain, and it is suppressed in the film with compressive strain. This clearly indicates the entanglement of the strain and the oxygen deficiency, that is, oxygen vacancies are induced to accommodate tensile strain and compressive strain suppresses the formation of oxygen vacancies. Even in the transport and magnetic measurements, the same tendency was observed. Tensile-strained LCMO film on STO substrate grown at 0.27 mbar shows suppressed $T_C$ and high residual resistivity because of the induced oxygen vacancies, while LCMO film on LSAT and LSAO substrates showed sharp transition with high $T_C$ and small $\rho_0$. Changing the oxygen pressure from 0.27 mbar to 0.2 mbar, $T_C$ and $\rho_0$ in compressive-strained LCMO film on LSAO substrate were stable compared to LCMO films on STO and LSAT substrates since the compressive strain suppresses formation of oxygen vacancies.

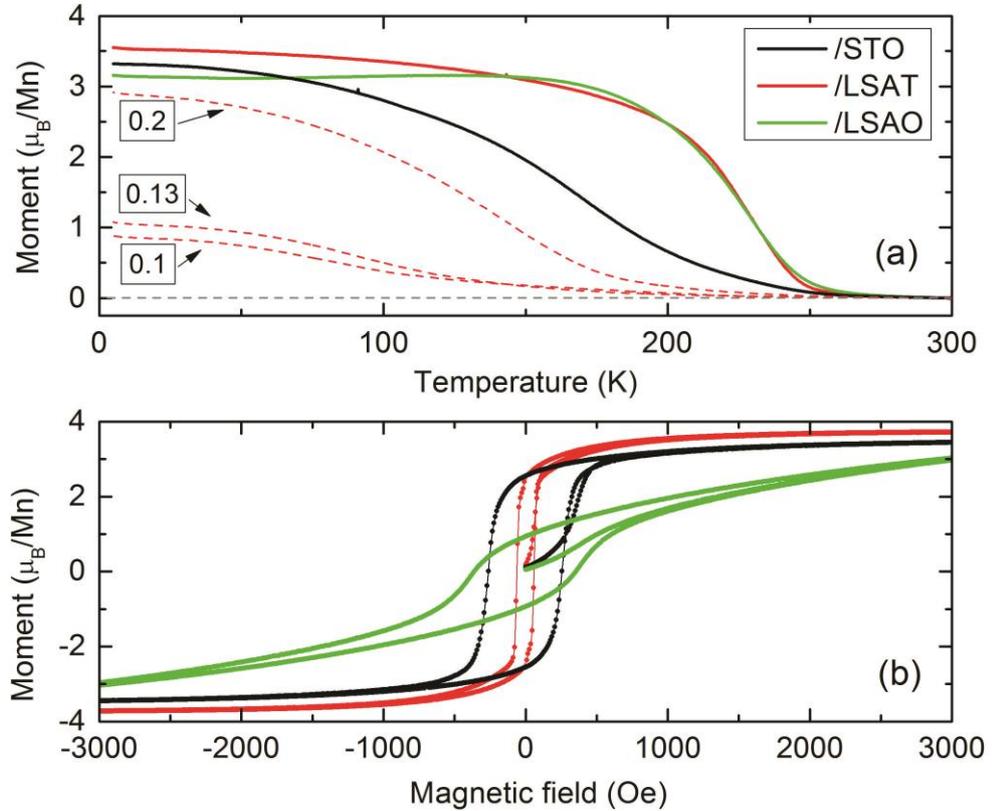

*Fig.7 Magnetic moment vs temperature, and hysteresis curves. (a) Magnetic moments during field cooling process under external field of 3000 (Oe). The bold curves indicate the magnetic moments of LCMO films grown at 0.27mbar on STO, LSAT and LSAO substrates. The broken lines show the moments of LCMO/LSAT grown at 0.2, 0.13 and 0.1mbar from the top line to the bottom one. (b) The magnetic loop at 5K on the LCMO samples grown at 0.27mbar.*

**Acknowledgments**

We are indebted to D. Samuelis for the critical reading of the manuscript and thank B. Stuhlhofer and Y. Link for expert technical support. K.K. owes a debt of gratitude to the Max-Planck Society for granting financial support.